\pdfoutput=1
\documentclass[11pt]{article}

\usepackage[a4paper,margin=1in]{geometry}
\usepackage[utf8]{inputenc}
\usepackage[T1]{fontenc}
\usepackage{libertinus}
\usepackage{microtype}

\usepackage{amsmath,amssymb,amsthm}
\usepackage{mathtools}
\usepackage{bm}

\usepackage{graphicx}
\usepackage{booktabs}
\usepackage{multirow}
\usepackage{array}

\usepackage{listings}

\usepackage{tikz}
\usetikzlibrary{arrows.meta, positioning, calc, fit, shapes.geometric}

\usepackage[numbers,sort&compress]{natbib}
\usepackage{xcolor}
\usepackage[colorlinks=true,linkcolor=blue!50!black,citecolor=blue!50!black,urlcolor=blue!50!black]{hyperref}
\usepackage[capitalize,noabbrev]{cleveref}

\newcommand{\vect}[1]{\bm{#1}}
\newcommand{\mat}[1]{\bm{#1}}
\newcommand{\R}{\mathbb{R}}

\newcommand{\layout}{\mathcal{L}}                      
\newcommand{\dctbasis}{\mat{\Phi}}                     
\newcommand{\lambdat}{\lambda_t}                       

\lstdefinestyle{paperpython}{%
  basicstyle=\ttfamily\small,
  keywordstyle=\color{blue!60!black}\bfseries,
  commentstyle=\color{gray!80!black}\itshape,
  stringstyle=\color{green!40!black},
  language=Python,
  breaklines=true,
  showstringspaces=false,
  numbers=none,
  frame=single,
  framerule=0.4pt,
  rulecolor=\color{gray!40},
}
\usepackage{fancyhdr}
\usepackage{tcolorbox}
\tcbuselibrary{skins,breakable}

\definecolor{futoblue}{HTML}{1F4E79}
\definecolor{futoborder}{HTML}{1F4E79}
\definecolor{futolightbg}{HTML}{F4F7FB}

\makeatletter

\let\@reportheader\@empty
\newcommand{\reportheader}[1]{\def\@reportheader{#1}}

\fancypagestyle{reportpages}{%
  \fancyhf{}%
  \fancyhead[C]{\fontsize{9pt}{11pt}\selectfont\@reportheader}%
  \fancyfoot[C]{\thepage}%
}
\fancypagestyle{reportfirstpage}{%
  \fancyhf{}%
  \fancyfoot[C]{\thepage}%
}

\AtBeginDocument{%
  \pagestyle{reportpages}%
  \thispagestyle{reportfirstpage}%
}

\def\@reportmetalist{}
\newcommand{\@reportmetaappend}[1]{\g@addto@macro\@reportmetalist{#1}}

\newcommand{\reportdate}[1]{%
  \@reportmetaappend{{\small\textbf{\color{futoblue}Date:}~#1\par}}%
}
\newcommand{\reportgithub}[1]{%
  \@reportmetaappend{{\small\textbf{\color{futoblue}GitHub:}~{\color{futoblue}#1}\par}}%
}
\newcommand{\reporthuggingface}[2][Hugging Face]{%
  \@reportmetaappend{{\small\textbf{\color{futoblue}#1:}~{\color{futoblue}#2}\par}}%
}
\newcommand{\reportcorrespondence}[1]{%
  \@reportmetaappend{{\small\textbf{\color{futoblue}Correspondence:}~#1\par}}%
}

\renewenvironment{abstract}{%
  \begin{tcolorbox}[
    enhanced,
    colback=futolightbg,
    colframe=futoborder,
    arc=10pt,
    boxrule=0.8pt,
    left=12pt, right=12pt,
    top=8pt, bottom=8pt,
    breakable,
    drop shadow={futoblue!8},
  ]%
  \centerline{\large\bfseries\color{futoblue} Abstract}%
  \vspace{0.4ex}%
}{%
  \ifx\@reportmetalist\@empty\else
    \par\vskip 1ex
    {\color{futoblue!25}\hrule height 0.4pt}%
    \vskip 1.2ex
    \@reportmetalist
  \fi
  \end{tcolorbox}%
  \vskip 0.5ex%
}

\makeatother

\title{\textbf{FUTO Swipe: Layout-Agnostic Neural Swipe Decoding}}

\author{%
  \textbf{David Lee Miller}\\
  FUTO
  \and
  \textbf{Aleksandras Kostarevas}\\
  FUTO
}

\date{}

\reportheader{FUTO Swipe Technical Report}
\reportdate{June 23, 2026}
\reporthuggingface[Models]{\url{https://huggingface.co/futo-org/futo-swipe}}
\reporthuggingface[Dataset]{\url{https://huggingface.co/datasets/futo-org/swipe.futo.org}}
\reportcorrespondence{\texttt{\{lee, alex\}@futo.tech}}

\begin{document}
\maketitle

\begin{abstract}
Neural swipe decoders are typically tied to the keyboard they
were trained on, requiring a new corpus and training run for
each layout. In this report, we document our approach toward training
models that can function on any contiguous mobile keyboard layout.
At each point along the swipe, our encoder
predicts whether the user is indicating a character and where on
the keyboard that character lies. The keyboard layout is
supplied at inference time and used to map the spatial and temporal prediction to
a logit at each key, rather than being learned during training.

Training neural models requires substantial data, but
public swipe data is limited, particularly for non-QWERTY layouts. We release
\texttt{swipe.futo.org}~\citep{futoSwipeDataset2026}, the
largest MIT-licensed swipe corpus we are aware of, containing over
1M donated swipes from more than 12k donor sessions.
To generalize beyond the English QWERTY layout, we apply
geometric augmentations to both the swipe trajectory and the
keyboard layout at every training step, forcing the model to
make predictions based on characteristics of the swipe gesture
rather than the training layout. The model generalizes to
layouts absent from training, in some cases more accurately than
the layout it was trained on. This combines the
layout-flexibility of an algorithmic decoder with the accuracy
of a neural model. Trained models are publicly
available.
\end{abstract}

\section{Introduction}
\label{sec:intro}

Swipe gesture typing on touchscreen keyboards is a popular
text-entry method on mobile devices. A swipe decoder maps the
continuous touch trajectory to a word in the user's lexicon, and
two decades of work have approached this mapping in several ways.
Original algorithmic work was based on template-matching
systems~\citep{kristensson2004shark2}, later joined by neural
CTC~\citep{graves2006ctc} decoders. Recurrent networks trained on English
QWERTY~\citep{alsharif2015gboardlstm} and
language-specific neural models for non-Latin
scripts~\citep{biju2020indicswipe} are now the most common
approach. Neural decoders require significant training data,
while the public availability of data is limited.
\citet{leiva2021howweswipe} released the largest open
English swipe corpus we are aware of, while production deployments at
Gboard~\citep{gboardNSS2024}, Apple iOS~\citep{mehra2020gan},
Microsoft SwiftKey, and Grammarly~\citep{grammarly2024swipe}
train on internal corpora that are not redistributed.

Template-matching decoders such as
SHARK\textsuperscript{2}~\citep{kristensson2004shark2} aim to make
swipe decoding usable on any keyboard by scoring a swipe
trajectory against an ideal template path through each word's key
sequence. Because the templates are regenerated from whichever
layout is active, the matching algorithm itself imposes no
constraint on the keyboard, but it struggles to disambiguate
words whose templates trace similar paths.
Popular keyboard layouts like QWERTY contain many confusable swipe
patterns. Consequently, layouts have been optimized for swipe-shape
distinctiveness~\citep{bi2016ijqwerty,smith2015optimizing,clearflowkeyboard}.

Neural decoders improve accuracy, but standard approaches
tie the resulting model to the keyboard it was trained on. The
layout enters as an input feature read by the
decoder~\citep{alsharif2015gboardlstm}, through a spatial
discretization keyed to specific key
positions~\citep{shen2024gesture2text}, or as separate sets of
parameters trained for each
layout~\citep{biju2020indicswipe}, and in each case the
resulting model cannot be applied to a different layout without
retraining.
Subsequent Gboard work uses finite-state transducer
composition~\citep{ouyang2017mobile,hellsten2017fstdecoding} over
a fixed layout, and the reported evaluations across this lineage
are within the model's training-set layout. In the
published literature a decoder is either layout-flexible
at template-matching accuracy, or competitive in accuracy at the
cost of being tied to a single layout. Closed production
deployments may avoid this trade-off, but no published work
addresses it directly.

The data needed to train a neural decoder is similarly
limited. The \emph{How We Swipe}
corpus~\citep{leiva2021howweswipe} was collected as a remote
web-based study and contains $109{,}338$ swipes from
$1{,}338$ users. The closed production deployments cited above
use private corpora whose scale and composition are not reported.
Layouts engineered to reduce swipe
ambiguity~\citep{smith2015optimizing,clearflowkeyboard} have no
publicly released swipe data at all. The limited availability of data
reinforces the fixed-layout deployment pattern, since a layout
without a public corpus also has no released neural decoder.

Augmentation is the standard approach to extending a training
corpus for improved generalization. Reported pipelines on swipe
trajectories include affine and time-scaling
transformations~\citep{grammarly2024swipe} and GAN-based synthetic
trajectory generation~\citep{mehra2020gan,chu2023wordgesturegan}.
Related work on short-stroke
gestures and on online or offline handwriting follows the same
template, varying the recorded trajectory while leaving the
reference target
unchanged~\citep{maslych2023strokeaug,hamdi2021geofreqbeta,wigington2017dataaug}.
They are effective at reducing the number of
recorded trajectories needed to fit a given keyboard, but they do
not address layout flexibility, so the fixed-layout deployment
pattern described above persists.

In this report, we introduce FUTO Swipe models, which prioritize
keyboard layout flexibility, on-device performance and decoding accuracy.
We demonstrate that two coordinated changes can combine the
layout-flexibility of an algorithmic decoder with the accuracy of
a neural model. First, in our model, the encoder consumes
the keyboard layout at inference as a tensor of $(x, y)$
coordinates for each key, and the spatial output head reads those
coordinates through a basis supplied at runtime, rather than
learning a separate parameter for each key. Second, every geometric augmentation applied at
training time is applied jointly to the trajectory and the
layout-key tensor, analogous to image and bounding-box
co-augmentation in vision~\citep{albumentations2020}. Both
choices are ablated in \Cref{sec:ablations}, and the empirical
evaluation in \Cref{sec:experiments} tests the resulting encoder
on real user swipes from a layout absent from the training data.

Finally, to address the open-data gap described above, we release
\texttt{swipe.futo.org}~\citep{futoSwipeDataset2026} alongside the
trained model. The corpus is an MIT-licensed collection of donated swipes
assembled by ongoing volunteer contribution.
Interested readers are encouraged to participate and
contribute.\footnote{\url{https://swipe.futo.org/}}

\section{Method}
\label{sec:method}

\begin{figure}[t]
  \centering
  \includegraphics[width=\linewidth]{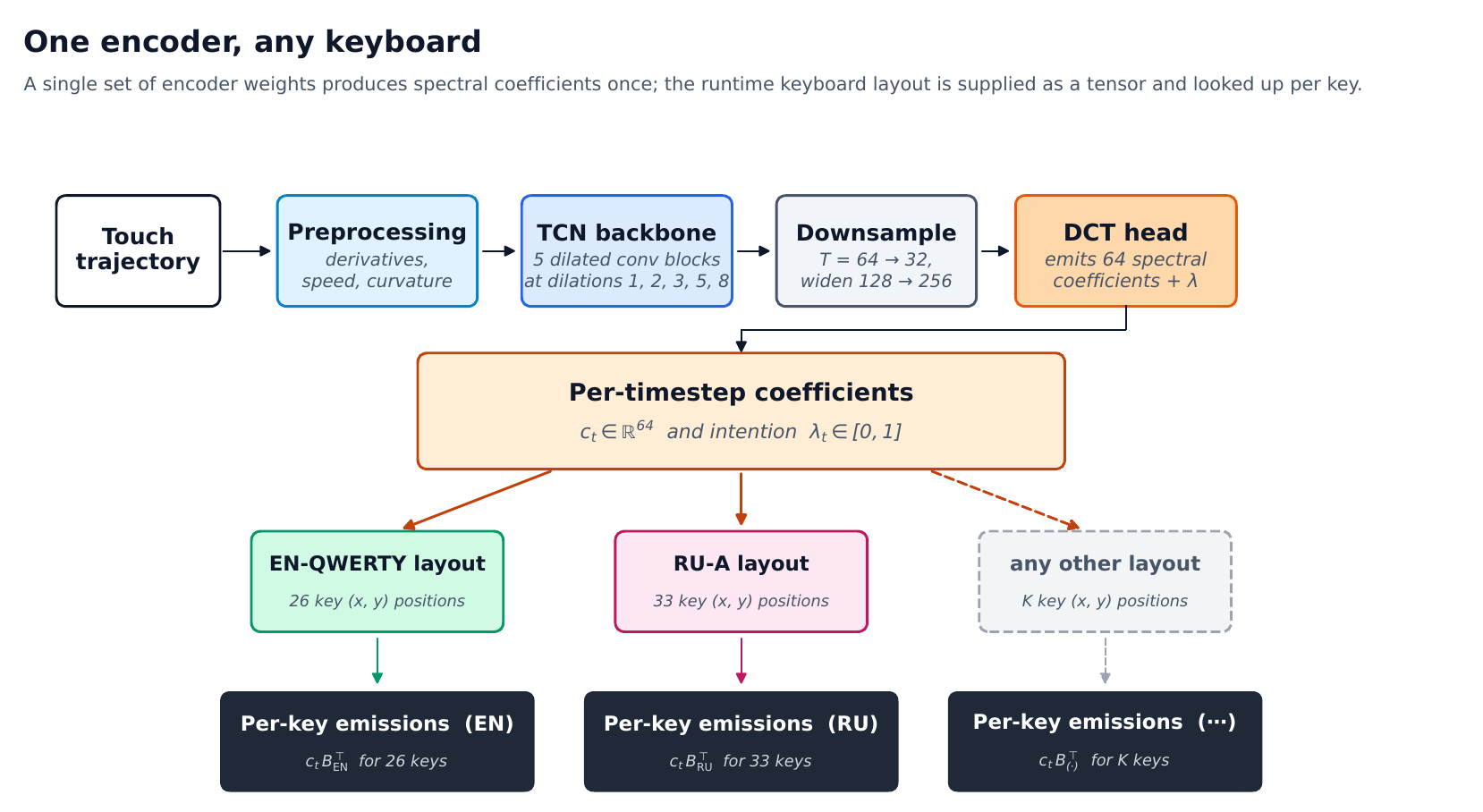}
  \caption{The encoder takes the trajectory and emits spectral
  coefficients $\vect{c}_t$ and intention scalar $\lambdat$ at
  each timestep. At inference, the runtime keyboard's key
  $(x, y)$ coordinates parameterize the fixed cosine basis
  $\dctbasis$, which is then sampled at those coordinates to
  obtain a logit at each key. The optional fixed-layout decoder
  (\Cref{ssec:decoder}, not shown) consumes the same shared
  features.}
  \label{fig:overview}
\end{figure}

Our model has two components: an encoder and an optional
fixed-layout decoder. The encoder consists of a trajectory-only
TCN backbone (\Cref{ssec:dct}) trained with coordinated trajectory
and layout-key augmentation (\Cref{ssec:augmentation}). At
inference, the keyboard enters the forward pass as a runtime
tensor of key coordinates. The optional fixed-layout DFSMN decoder
over frozen encoder features (\Cref{ssec:decoder}) refines
accuracy where layout-specific training data is available.
\Cref{fig:overview} shows the encoder pipeline.

\subsection{Layout-agnostic encoder via a spectral spatial head}
\label{ssec:dct}

\paragraph{Output heads}
Two predictions are emitted at each timestep by independent
linear projections of the backbone hidden state: a
scalar \emph{intention} $\lambdat \in [0,1]$ marking the points
along the gesture at which the user indicates a character, and a
64-coefficient spectral pattern $\vect{c}_t \in \R^{64}$ that
locates the intention on the keyboard.
The spatial pattern is layout-conditioned only at
evaluation time, through the basis defined in
\Cref{eq:dct-basis}. The CTC emission
distribution is recovered by a factorized softmax that emits blank
from $1-\lambdat$ and characters from $\sigma(\vect{c}_t\,
\dctbasis^{\top})\cdot \lambdat$ (full derivation and sweep against
the joint $(K{+}1)$-way softmax baseline in \Cref{app:abl-lambda}).

\paragraph{DCT formulation}
Let $N$ denote the head's spatial resolution (with $N{=}8$ in
production, so $N^2 = 64$ coefficients, and the spatial-head
ablation of \Cref{ssec:abl-spatial-head} sweeping other values). We index the
$N^2$ coefficients as $c_{t,(u,v)}$ for $u, v \in \{0, \dots, N{-}1\}$
and treat them as the coefficients of a 2D separable cosine basis
over $[0,1]^2$. Let $K$ denote the number of keys in the active
layout. Given the key-center coordinates of the active layout
$\layout = \{(u_k, v_k)\}_{k=1}^{K}$, normalized to $[0,1]^2$, we
construct a fixed basis matrix $\dctbasis \in \R^{K \times N^2}$
whose row for each key holds the cosine basis evaluated at that
key's coordinates:
\begin{equation}
  \dctbasis[k,\, (u,v)]
  \;=\;
  \cos(\pi\, u\, u_k) \cdot \cos(\pi\, v\, v_k),
  \quad u, v \in \{0, \dots, N{-}1\}.
  \label{eq:dct-basis}
\end{equation}
The basis is computed once per layout and reused across timesteps. Key logits at timestep $t$ are the inner
product between the emitted coefficients and each key's row of the
basis,
\begin{equation}
  z_{t,k}
  \;=\;
  \sum_{u, v = 0}^{N-1}
  c_{t,(u,v)} \cdot \cos(\pi\, u\, u_k)\, \cos(\pi\, v\, v_k)
  \;=\;
  \vect{c}_t \cdot \dctbasis[k],
  \label{eq:dct-lookup}
\end{equation}
or in matrix form, $\vect{z}_t = \vect{c}_t\, \dctbasis^{\top} \in
\R^{K}$. Geometrically, the coefficients $\vect{c}_t$ pick a 2D
spatial pattern on the unit square at each timestep, and $z_{t,k}$
is the value of that pattern sampled at key $k$'s position. The
spectral pattern emitted by the encoder corresponds to intended
character selections at spatial and temporal locations along the
user's gesture. Although this intention cannot be measured
directly, the model learns to predict it from characteristics of
the gesture rather than from the underlying layout. This
abstraction yields a representation that generalizes to layouts
absent from training.

\begin{figure}[t]
  \centering
  \includegraphics[width=\linewidth]{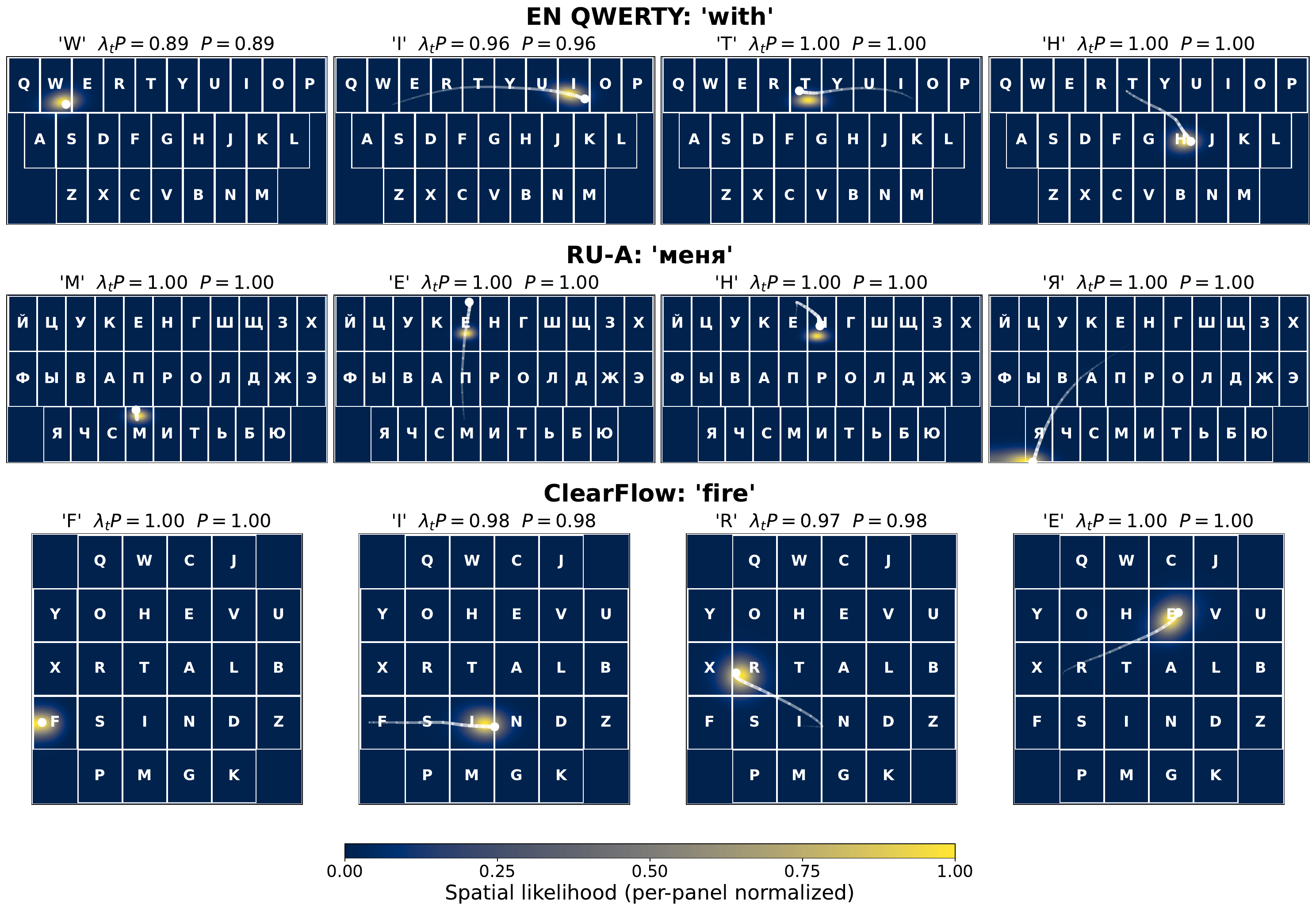}
  \caption{Spatial likelihood field of the encoder on
    three val swipes.
    Top: English-QWERTY \emph{with}.
    Middle: Russian-JCUKEN \emph{menya} on held-out RU-A.
    Bottom: ClearFlow~\citep{smith2015optimizing,%
    clearflowkeyboard} \emph{fire} on a layout never seen at
    training time. Each panel renders the key-logit field
    $\vect{c}_t\, \dctbasis^{\top}$ at the timestep $t^{*}$ that
    maximizes $\lambdat \, P(\text{ch} \mid t)$ for the column's
    character. White trail: trajectory leading into $t^{*}$.
    Per-panel titles report $\lambdat \, P(\text{ch})$ and
    $P(\text{ch})$ at $t = t^{*}$.}
  \label{fig:demo-spatial-field}
\end{figure}

\paragraph{Design}
The encoder backbone is a 1D temporal convolutional network. Each
block applies a dilated depthwise convolution, batch normalization,
a $1{\times}1$ expansion with a gated linear unit, a global
response normalization~\citep{woo2023convnextv2}, a $1{\times}1$
projection back to the trunk width, and a squeeze-and-excitation
gate~\citep{hu2018se} before the residual sum, following the
ConvNeXt~\citep{liu2022convnext, woo2023convnextv2} block adapted
to one dimension. The deployed encoder stacks five blocks with
dilations $\{1, 2, 3, 5, 8\}$ at trunk width $128$ and expansion
factor $4$. A 2$\times$ adapter (stride-2, kernel-size-2 1-D
convolution + batch norm) halves the time axis and widens the
hidden state to the spatial head: $T_{\text{in}} = 64 \to
T_{\text{out}} = 32$.

\paragraph{Input features}
Raw $(x, y, t)$ point streams are resampled to 60\,Hz (the modal
sampling rate in the dataset) and then to $T_{\text{in}} = 64$
evenly-spaced points by linear interpolation. From the resampled
$(x, y)$ stream we derive an 8D timestep feature vector via a
fixed Savitzky--Golay filter (7-tap, polynomial order 2): position
$(x, y)$, velocity $(\dot x, \dot y)$, acceleration
$(\ddot x, \ddot y)$, speed $\sqrt{\dot x^2 + \dot y^2}$, and
curvature (the rate of change of $\mathrm{atan2}(\dot y, \dot x)$,
clamped to $[-2, 2]$). The filter is implemented in torch and
exported as part of the model graph.

\paragraph{Implementation}
Both linear projections in the head are zero-initialized, so at
step zero every $z_{t,k}$ is identically zero and $\lambdat = 0.5$
uniformly. $\dctbasis$ is computed once per layout from the runtime key
coordinates (\Cref{eq:dct-basis}) and cached until the layout
changes, and key logits
are a single batched matrix multiplication $\vect{z}_t =
\vect{c}_t\, \dctbasis^{\top}$. Tensor shapes for every stage of
the forward pass are tabulated in \Cref{app:mobile-deployment}.

\paragraph{Training}
The encoder is trained with AdamW~\citep{loshchilov2019adamw} for
$120$ epochs at batch size $1024$ on the corpora of
\Cref{ssec:setup}, under the augmentation pipeline of
\Cref{ssec:augmentation}. The training loss is CTC plus an
emission-count regularizer that stabilizes the gate $\lambdat$
against the peakiness of standard CTC blank emission (full
derivation, ablation, and comparison against the joint
$(K{+}1)$-way softmax in \Cref{app:abl-lambda}). Full optimizer
schedule and regularization values are in
\Cref{app:training-details}.

\subsection{Coordinated trajectory and layout-key augmentation}
\label{ssec:augmentation}

The augmentation pipeline runs each batch on the GPU and applies seven
stages in order: y-scale, x-scale, shear, flips, rotation,
translation, and time reversal. The first six geometric stages are
applied identically to the trajectory tensor and to the
training-time layout-key tensor, so the augmented keyboard remains
geometrically consistent with the augmented swipe. Time reversal
reverses both the temporal axis of the trajectory and the target
word so the CTC label sequence stays aligned. At inference the
runtime layout-key tensor carries only key centroids
(\Cref{ssec:dct}). Parameter ranges for each stage and domain-specific
rules (Indic-aware y-scale skip, in-bounds rotation rejection) are
in \Cref{app:augmentation-stages}. The ablation in
\Cref{ssec:abl-augmentation} quantifies the contribution of each
stage to cross-layout transfer.

\subsection{Optional fixed-layout decoder}
\label{ssec:decoder}

\paragraph{Design}
For a target layout with training data, we precompute the encoder's
output at each timestep over the training set ($K{+}1$-way log
emissions, $64$ DCT coefficients, scalar intention $\lambdat$) and
train a small DFSMN-style
network~\citep{zhang2018dfsmn} using those input features. The decoder's CTC
head is zero-initialized and its logit output is added to the
encoder's log emissions via a residual skip.
The decoder is therefore a correction over the encoder baseline,
specialized to the target layout and language.

Joint fine-tuning would force a distinct encoder for each
$\langle$layout, language$\rangle$ pair. We freeze the encoder
instead, reusing one set of weights across all layouts. Layouts
without sufficient training data fall back to encoder-only beam
search.

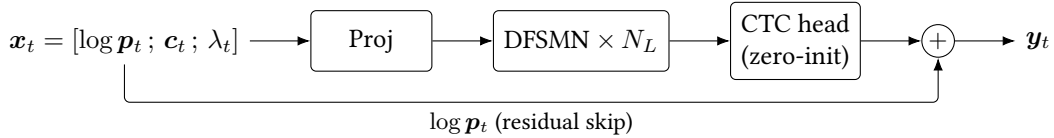
\begin{figure}[h]
\centering
\begin{tikzpicture}[
  font=\small,
  >=Latex,
  node distance=10mm and 8mm,
  block/.style={draw, rounded corners=2pt, align=center,
                minimum height=8mm, minimum width=16mm,
                inner sep=4pt},
  io/.style={align=center, font=\small},
  add/.style={draw, circle, inner sep=1pt, minimum size=4mm, font=\footnotesize},
]
  \node[io] (in) {$\vect{x}_t = [\log \vect{p}_t \,;\, \vect{c}_t \,;\, \lambdat]$};
  \node[block, right=of in] (proj) {Proj};
  \node[block, right=of proj] (dfsmn) {DFSMN $\times\, N_L$};
  \node[block, right=of dfsmn] (head) {CTC head\\(zero-init)};
  \node[add, right=of head] (sum) {$+$};
  \node[io, right=of sum] (out) {$\vect{y}_t$};

  \draw[->] (in) -- (proj);
  \draw[->] (proj) -- (dfsmn);
  \draw[->] (dfsmn) -- (head);
  \draw[->] (head) -- (sum);
  \draw[->] (sum) -- (out);

  \coordinate (skipstart) at ($(in.south) + (0, -5mm)$);
  \coordinate (skipend) at (sum |- skipstart);
  \draw[->, rounded corners=2pt]
    (in.south) -- (skipstart)
    -- node[below, font=\footnotesize] {$\log \vect{p}_t$ (residual skip)} (skipend)
    -- (sum.south);
\end{tikzpicture}
\caption{Fixed-layout decoder. The encoder feature
  $\vect{x}_t = [\log \vect{p}_t \,;\, \vect{c}_t \,;\, \lambdat]$
  is projected to the hidden width $H_d$, passed through $N_L = 8$
  DFSMN blocks (each with an internal bottleneck of width $P_d$),
  and mapped to per-character logits by a zero-initialized CTC head.
  The original log-emissions $\log \vect{p}_t$ are added back via a
  residual skip, so the decoder begins training as the identity on
  the encoder output and learns a correction over it.}
\label{fig:decoder-skip}
\end{figure}

\paragraph{Formulation}
Let $\vect{x}_t = [\log \vect{p}_t \,;\, \vect{c}_t \,;\, \lambdat]
\in \R^{(K{+}1) + 64 + 1}$ be the frozen encoder feature at timestep
$t$. The decoder pipeline (\Cref{fig:decoder-skip}) projects
$\vect{x}_t$ to the hidden width $H_d$, applies a stack of $N_L$
DFSMN blocks in sequence, passes the result through a zero-initialized
CTC head, and adds the original encoder log-emissions $\log
\vect{p}_t$ back via a residual skip. The CTC-head weight and bias
are zero-initialized, so $\vect{y}_t \!\equiv\! \log \vect{p}_t$ at
step zero and the decoder begins training as the identity on the
encoder output.

\paragraph{Implementation}
The decoder uses $N_L{=}8$ DFSMN blocks with hidden width
$H_d{=}256$, bottleneck projection $P_d{=}64$, and a symmetric
memory context of $7$ frames (length-$15$ depthwise kernel) on the
bottleneck axis. With the $(K{+}1)=27$-dim log-emission slice, $64$
DCT coefficients, and $1$-dim $\lambdat$, the input is $92$
dimensional.

Encoder features are precomputed and reused across all decoder
hyperparameter sweeps (no augmentation is applied at this stage).
The data loader reconstitutes the $K{+}1$-way log emissions on the
fly by evaluating the layout's basis and applying
\Cref{eq:lambda-emission}.

\paragraph{Ranking loss}
Beam-search decoding over a lexicon trie is a ranking task. At
evaluation we read the $K$-best beams and choose the highest-scoring
word under a length- and frequency-aware combination of CTC cost
and lexical priors (\Cref{ssec:setup}). We add a pairwise ranking
objective in the LambdaLoss
family~\citep{wang2018lambdaloss,burges2010lambdamart}. The
ground truth and a pool of mined hard negatives are scored with
the same length-normalized CTC score used at inference,
\begin{equation}
  s(w \mid \vect{y})
  \;=\;
  -\,\frac{\mathrm{CTC}(w \mid \vect{y})}{L_w^{\gamma}}
  \,+\, \lambda_{\text{f}}\,\log f_w
  \,+\, \beta\, L_w,
  \label{eq:scoring}
\end{equation}
where $\mathrm{CTC}(w \mid \vect{y})$ is the CTC negative
log-likelihood (a cost, so the leading minus turns it into a
score), $L_w$ is the word length, and $f_w$ is its corpus frequency. The
three scoring parameters $(\gamma, \lambda_{\text{f}}, \beta)$ are
trained jointly with the DFSMN by a separate SGD optimizer. We
freeze $\gamma$ at $0.30$ to keep it from oscillating against the
linear length term and clamp $\lambda_{\text{f}} \ge 0$ after each
step. The pairwise loss is the NDCGLoss2++
variant of~\citet{wang2018lambdaloss}, summing
$\mathrm{softplus}(-\sigma\,(s_{\text{gt}} - s_{\text{neg}}))$ over
$(\text{gt}, \text{neg}_k)$ pairs weighted by
$\mu\!\cdot\!\Delta\mathrm{NDCG2} + \Delta\mathrm{LambdaRank}$,
with $\mu{=}10$ and $\sigma{=}1$.

The hard-negative pool is mined offline for each layout from a
contrastively trained 128-dim trajectory
embedding~\citep{swipealot2025}: for each target word,
$k$-NN queries are aggregated by hit count across the $S_w$ swipe
samples of that word, and the most-frequently-retrieved words
form a pool of up to $128$ hard negatives. The pool
is locked at training start and batch-random subsampling
provides stochasticity. The English pool is released as an open
dataset~\citep{futoSwipeNegatives2026}. Out-of-corpus words fall
back to the embedder's text-only encoding path. We gate the
ranking term with a validation-CTC threshold: no gradient until
$\mathrm{val\_ctc} < 0.205$, after which the term is unlocked for
the remainder of training.

\paragraph{Consistency-regularized CTC}
We add CR-CTC~\citep{yao2025crctc} as a consistency regularizer
over two noised views of the encoder features. With
$\vect{x}_t^{(a)}, \vect{x}_t^{(b)} = \vect{x}_t +
\boldsymbol\epsilon^{(a,b)}$ and $\boldsymbol\epsilon \sim
\mathcal{N}(0, \sigma^2 I)$ at $\sigma = 0.10$, both views are
passed through the decoder and the loss enforces a forward KL
between the two output distributions
\begin{equation}
  \mathcal{L}_{\text{CR}}
  \;=\;
  \mathrm{KL}\!\bigl(\,
    \mathrm{softmax}(\vect{y}_t^{(a)})
    \,\Vert\,
    \mathrm{softmax}(\vect{y}_t^{(b)})
  \bigr).
  \label{eq:cr-ctc}
\end{equation}
Among the alternatives we tried (dropout-only consistency in the
spirit of SimCSE~\citep{gao2021simcse}, time masking, channel
masking), additive Gaussian jitter on the input features was the
most effective. The top-3 accuracy gain at the picked
checkpoint is small, but importantly CR-CTC stabilizes training against
overfitting over a longer schedule (\Cref{ssec:abl-crctc}).

\paragraph{Loss construction}
The combined loss is
\begin{equation}
  \mathcal{L}
  \;=\;
  0.3 \cdot \mathcal{L}_{\text{CTC}}
  \;+\; 5.0 \cdot \mathcal{L}_{\text{rank}}
  \;+\; 0.1 \cdot \mathcal{L}_{\text{CR}},
  \label{eq:decoder-loss}
\end{equation}
with $\mathcal{L}_{\text{rank}}$ active only after the
validation-CTC gate has opened. Decoder weights are trained with
Lion~\citep{chen2023lion} and the scoring-head parameters with a
separate SGD optimizer. An exponential moving average (EMA) copy of the
decoder is used for validation, beam-search evaluation, and the
exported checkpoint. Full optimizer schedule and regularization
values are in \Cref{app:training-details}.

\section{The \texttt{swipe.futo.org} Dataset}
\label{sec:dataset}

We introduce \texttt{swipe.futo.org}~\citep{futoSwipeDataset2026},
an MIT-licensed swipe-typing corpus (primarily QWERTY) collected
by volunteer donation. This section documents the collection
methodology, the released schema, the filtering pipeline, and
known limitations.

\subsection{Motivation and scope}
\label{ssec:dataset-motivation}

Open swipe data is limited to a small number of releases.
\emph{How We Swipe}~\citep{leiva2021howweswipe} is a fixed
16-sentence remote web-based study. Other published swipe-decoder
work trains against private production corpora. Our collection method
follows the same overall design as \emph{How We Swipe}: a
web-rendered virtual QWERTY with no live decoding, and
sentence-based transcription with word-by-word visual feedback.
Stimuli are drawn from Mozilla Common
Voice~\citep{ardila2020commonvoice}.
Collection is ongoing volunteer donation rather than a fixed
test, so the released data is actively being updated with
additional subsets. The corpus is released under the MIT license.

\subsection{Collection methodology}
\label{ssec:dataset-collection}

Volunteers visit \url{https://swipe.futo.org} on a touchscreen
mobile device. The site does not render on desktop browsers
(detected by user agent and viewport width), so all collected
swipes originate from real touch hardware. A user is assigned a
single short-lived session id with no link to the donor's
identity. After accepting an on-screen consent and instruction
screen, the donor is shown a randomly chosen sentence in
word-by-word context. Each word is highlighted in turn and the
donor swipes that word on the QWERTY keyboard rendered below the
prompt. A \emph{Skip} button advances past the current word and
writes a sentinel record. A \emph{Del} button steps back so the
donor can retry. Retries upsert on \textsf{(session, sentence,
word)} and the latest attempt is what we release.

Touch input is captured at the device's native rate ($60$--$120$\,Hz). Each event
records normalized $(x, y)$ touch coordinates and a millisecond
timestamp $t$. A saved record contains the point
sequence $\{(x_i, y_i, t_i)\}_{i=1}^{T}$, canvas width and height in
pixels, the device orientation reported by the browser, the
challenge word, and the sentence and word indices.

\subsection{Stimulus material}
\label{ssec:dataset-stimulus}

For the primary \texttt{swipe-1} subset, stimuli are drawn
uniformly at random from the English sentence pool of Mozilla
Common Voice~\citep{ardila2020commonvoice}, which sources its
sentences from Wikipedia article text and contributes
approximately $1.3$\,M sentences to our prompt pool. Sentences
are typically short factual statements, which skews the released
word-frequency distribution toward written encyclopedic English
(proper nouns, place names, named entities). Subsequent runs
draw on different stimulus pools
(\Cref{ssec:dataset-followups}).

\subsection{Filtering}
\label{ssec:dataset-filtering}
The released snapshot is filtered. Swipes that fail basic
structural validity checks (degenerate trajectory length,
non-monotonic timestamps, out-of-bounds coordinates, implausible
duration, or mismatch between challenge word and recorded word) are
dropped, as are explicit \emph{Skip} sentinels. Approximately $5\%$
of submissions are removed by these filters.

For experiments in this paper we additionally drop swipes that do
not visibly follow the target word's keys. About $0.4\%$ of swipes
are removed by this check. The filter is not applied to the
released dataset.

\subsection{Statistics}
\label{ssec:dataset-stats}

\begin{table}[h]
\centering
\caption{Descriptive statistics of the \texttt{swipe-1} subset of the
released \texttt{swipe.futo.org} dataset.}
\label{tab:dataset-stats}
\begin{tabular}{lr}
\toprule
\multicolumn{1}{c}{Quantity} & \multicolumn{1}{c}{Value} \\
\midrule
Total swipes (filtered)            & 1{,}043{,}789 \\
Unique target words                & 108{,}759 \\
Unique anonymized sessions         & 12{,}278 \\
Unique sentences seen as stimulus  & 106{,}546 \\
Median swipe duration (ms)         & 799 \\
Median trajectory length (samples) & 55 \\
Filtering rate                     & $\sim$5\% \\
\bottomrule
\end{tabular}
\end{table}

\subsection{Splits}
\label{ssec:dataset-splits}
The primary \texttt{swipe-1} corpus is partitioned by donor
session into train, validation, and test splits of $939{,}550$,
$54{,}269$, and $49{,}970$ swipes respectively. All swipes from a session belong to the same split,
so val and test numbers reflect generalization to unseen donor sessions.
The vocabulary is not held out by construction.

\subsection{Subsequent collection runs}
\label{ssec:dataset-followups}

Four smaller collection runs have been released alongside the
main \texttt{swipe-1} corpus, adding roughly $175{,}000$ swipes total. Each
targets a specific gap: informal language (\texttt{swipe-2},
$28{,}095$ swipes), unique-word coverage (\texttt{swipe-3}, $38{,}228$),
confusable word sets (\texttt{swipe-4}, $50{,}300$), and additional layouts and languages
(\texttt{swipe-5}, $59{,}247$). The ClearFlow validation data used in
\Cref{sec:experiments} is drawn from \texttt{swipe-5}; the remaining ten
layouts there are smaller (under $3{,}000$ swipes each) and are
not used for evaluation in this paper. Subsequent runs are
released unfiltered; the \texttt{distance} field is the
recommended filter.

\subsection{Limitations and biases}
\label{ssec:dataset-limitations}
\paragraph{Layout and language coverage}
The \texttt{swipe-1} release is English-QWERTY only. The subsequent runs
of \Cref{ssec:dataset-followups} broaden coverage to eleven
layouts and eight languages, but per-layout and per-language
counts outside English-QWERTY remain small at the time of writing.

\paragraph{Donor self-selection}
Donors are FUTO website visitors, skewed toward open-source and
privacy-conscious users. Demographic, handedness, and dominant-hand
distributions are not recorded.

\paragraph{Skip-induced sentence gaps}
Donors can skip individual words, so a sentence may appear with
holes. The schema preserves \texttt{sentence\_id} and
\texttt{word\_index} for reconstruction where context survives.

\paragraph{Stimulus register}
The Mozilla Common Voice
sentences~\citep{ardila2020commonvoice} are sourced from Wikipedia
article text, which skews the vocabulary toward formal
encyclopedic English with underrepresentation of conversational
idioms, slang, brand names, and other colloquial language.

\section{Experiments}
\label{sec:experiments}

\subsection{Setup}
\label{ssec:setup}

\paragraph{Evaluation corpora}
English numbers come from \texttt{swipe.futo.org} (\Cref{sec:dataset})
val and test splits. Russian
swipe validation results come from the Yandex Cup 2023 NeuroSwipe
data\footnote{Yandex Cup 2023, \url{https://yandex.com/cup/2023}.}.
Only the val split contains
ground truth labels. The Yandex corpus covers two Cyrillic JCUKEN
layouts: RU-A ($31$ keys, $9{,}416$ val samples after dropping
targets with under two swipeable characters) and RU-B ($32$ keys,
$584$ val samples). We combine them into a single row of size
$9{,}970$ after the trajectory-quality filter of
\Cref{ssec:dataset-filtering}. All layouts
are normalized to a $[0,1]^2$ frame. The
encoder is trained on English \texttt{swipe.futo.org} only.
Russian and ClearFlow are held out from training. Per-row
ablation encoders later in the paper are also trained on English
only. The language dependence of scoring is examined in
\Cref{tab:abl-scoring-language}.
ClearFlow~\citep{clearflowkeyboard} validation data comes from
$n{=}11{,}028$ swipes we collected on the ClearFlow layout,
released as part of \texttt{swipe-5} in the
\texttt{swipe.futo.org}
dataset~\citep{futoSwipeDataset2026}.

\paragraph{Decoding}
Beam search is trie-constrained with beam width $100$ and uses
length-aware beam pruning. The pruning score $s_{\text{prune}} =
s_{\text{ctc}} / \max(d,1)^{\gamma_\text{p}} + \beta_\text{p}
\cdot d$ (depth $d$) has coefficients tuned to maximize beam
recall@$K$.
The trie for each layout is the deployment lexicon (an
AOSP-format wordlist~\citep{futoAndroidKeyboard}: $162{,}185$
English entries, $220{,}500$ Russian entries) extended with the
evaluation target vocabulary, isolating spatial decoding from
OOV coverage. Candidates are
rescored by \Cref{eq:scoring}. Two inference modes are evaluated:
encoder-only beam search on the $\lambdat$-gated log-emissions of
\Cref{eq:lambda-emission}, and encoder-plus-decoder
beam search on the residual-skip output of
\Cref{fig:decoder-skip}.

\paragraph{Tuning}
The pruning coefficients $(\gamma_\text{p}, \beta_\text{p})$ and
the scoring coefficients $(\gamma, \lambda_{\text{f}}, \beta)$ of
\Cref{eq:scoring} are optimized in two stages, both on the English val
split (matching the encoder's training scope). Russian and
ClearFlow are held out from both stages. Pruning is tuned first,
using beam recall@$K$ as the optimization metric.
Scoring is tuned second on the surviving beams, using
an ensemble of metrics:
$\tfrac14(\text{top1} + \text{top3} + \text{mAP@5} +
\text{macroF1@5})$, where macroF1@5 and mAP@5 are word-level
metrics restricted to words with five or more examples in the
evaluation set.

Each stage uses Optuna with a tree-structured Parzen estimator
($50$ pruning trials, $3{,}000$ scoring trials). The frequency
term $\log f_w$ in \Cref{eq:scoring} reads the trie's stored
frequency field, which follows the AOSP wordlist convention $f =
\mathrm{round}\bigl(255 \cdot (\log_{10} f_w^{\text{raw}} -
\log_{\min}) / \log_{\text{range}}\bigr)$, a $0$--$255$ integer
proportional to log raw frequency. To avoid recall@$K$
drop from out-of-vocabulary targets, each tuning loop adds the
eval set's target words into the trie before decoding.
An ablation of the scoring formula is in
\Cref{app:abl-scoring-terms}. The language dependence of the
scoring tune itself is in \Cref{tab:abl-scoring-language}.
\Cref{tab:main-encoder-only,tab:main-with-decoder} share these
optima across layouts. Ablation tables
(\Cref{tab:abl-augmentation,tab:abl-spatial-head})
re-tune per row so each variant is evaluated under its own
optimum. \Cref{tab:abl-crctc} uses the decoder-mode shared
optimum.

Beyond the two stages tuned above, production deployment
optionally enables a context language model (LM) that adds an
$\alpha \cdot s_{\text{LM}}$ term to \Cref{eq:scoring}, where
$s_{\text{LM}}$ is the LM's log-likelihood of the candidate word
given preceding context. When enabled, scoring tuning extends to
four parameters. The LM is trained as a separate component and is
not evaluated in this paper.

\subsection{Main results}
\label{ssec:main-results}

\Cref{tab:main-encoder-only} reports decoding accuracy by layout.
Each block compares the SHARK\textsuperscript{2}
template-matching baseline against our encoder under two
scoring-tune scopes. The encoder itself is the same across both
scopes. Only the validation data used for the two-layer scoring tune
varies.

Our encoder's top-1 on ClearFlow exceeds its top-1 on the
in-domain QWERTY it was trained on, even though no ClearFlow
swipes appear in the training data.
SHARK\textsuperscript{2} also performs well on ClearFlow, since
the layout was optimized for template-matching shape uniqueness,
but our encoder leads it by roughly five points. The two other
rows, held-out JCUKEN and in-domain QWERTY, both show wider
encoder-over-SHARK\textsuperscript{2} margins.

The EN+RU tune scope shows that when a small amount of
language-specific validation data is available, retuning the
scoring on the \emph{same encoder} produces a higher top-$1$ on
the new language. Adding RU val to the scoring objective
(balanced with EN val) improves the held-out Russian top-$1$ by
$+2.79$ pt at a cost of $-0.75$ pt on the
in-domain QWERTY layout. ClearFlow is held
out from this tune as well and changes by less than half a point
on top-1.
\Cref{tab:abl-scoring-language} shows that the frequency term
does not transfer easily between languages, and that fitting
language-specific scoring parameters is preferable when the data
is available.

As expected, the ClearFlow optimization objective is well chosen for template
matching algorithms, and SHARK\textsuperscript{2} performs
well compared to the unoptimized layouts. Although our encoder
also performs well on ClearFlow, our modeling of
\emph{intention} is not matched to the shape uniqueness
objective that ClearFlow uses. We postulate that the improvement in
accuracy compared to QWERTY is primarily due to the increased number of
rows in the layout, which reduces colinearity of letter trigrams.
\Cref{ssec:kasroz} explores this hypothesis with a layout we
construct using the encoder itself as the cost function.

\begin{table}[t]
\centering
\caption{Decoding accuracy at beam width $100$, grouped by layout.
Scoring is tuned on English validation data alone (Tune = EN, with Russian and ClearFlow
held out), and scoring tuned on a balanced English + Russian
validation data (Tune = EN+RU, with ClearFlow held out).}
\label{tab:main-encoder-only}
\begin{tabular}{llllccc}
\toprule
Layout & $N$ & Method & Tune & Top-1 (\%) & Top-3 (\%) & Top-10 (\%) \\
\midrule
QWERTY    & $52{,}629$ & SHARK\textsuperscript{2} & EN      & $80.05$ & $90.47$ & $94.54$ \\
          &            & Ours                     & EN      & $92.94$ & $97.46$ & $98.60$ \\
          &            & Ours                     & EN+RU   & $92.19$ & $97.32$ & $98.57$ \\
\midrule
JCUKEN    & $9{,}970$  & SHARK\textsuperscript{2} & EN      & $59.31$ & $74.11$ & $82.70$ \\
          &            & Ours                     & EN      & $83.11$ & $91.26$ & $95.42$ \\
          &            & Ours                     & EN+RU   & $85.90$ & $93.01$ & $96.03$ \\
\midrule
ClearFlow & $11{,}028$ & SHARK\textsuperscript{2} & EN      & $92.18$ & $97.23$ & $98.33$ \\
          &            & Ours                     & EN      & $96.84$ & $98.98$ & $99.48$ \\
          &            & Ours                     & EN+RU   & $96.46$ & $99.08$ & $99.46$ \\
\bottomrule
\end{tabular}
\end{table}

\Cref{tab:main-with-decoder} reports encoder-only and
encoder-plus-decoder accuracy on the English val and test splits.
The fixed-layout decoder raises top-$1$ by $0.55$ pt (val) and
$0.76$ pt (test). Russian decoders are not trained in this work.

\begin{table}[t]
\centering
\caption{Optional fixed-layout decoder on the QWERTY
layout. Encoder-only rows use the encoder of
\Cref{tab:main-encoder-only}. Scoring and pruning
are tuned on EN validation separately.}
\label{tab:main-with-decoder}
\begin{tabular}{lcccc}
\toprule
Setting & Split & Top-1 (\%) & Top-3 (\%) & Top-10 (\%) \\
\midrule
Encoder only        & val   & 92.94 & 97.46 & 98.60 \\
Encoder + decoder   & val   & 93.49 & 97.85 & 99.08 \\
Encoder only        & test  & 92.54 & 97.33 & 98.54 \\
Encoder + decoder   & test  & 93.30 & 97.97 & 99.16 \\
\bottomrule
\end{tabular}
\end{table}

\subsection{KASROZ: a swipe-optimized layout for neural decoding}
\label{ssec:kasroz}

This subsection reports a small experiment that uses the encoder
itself as the cost function for designing a swipe-optimized
layout, then evaluates the resulting layout on real user swipes
collected against it. The layout-agnostic property of the encoder
allows it to be used to evaluate any layout, so optimization can
be performed directly with the model.

\begin{figure}[h]
  \centering
  \includegraphics[width=0.72\linewidth]{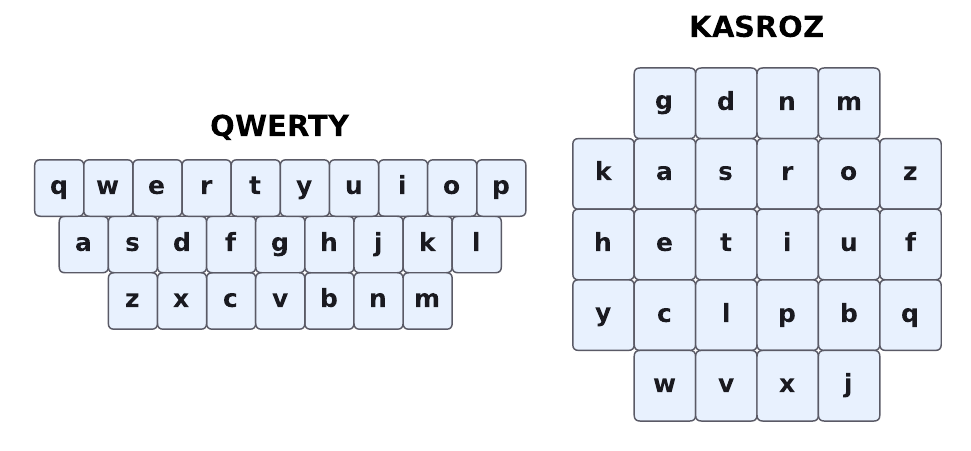}
  \caption{KASROZ keyboard layout (right) compared to QWERTY
  (left). KASROZ uses a $5$-row $4{-}6{-}6{-}6{-}4$ ortho grid. The name
  comes from the letter sequence in row $2$.}
  \label{fig:kasroz-layout}
\end{figure}

\paragraph{KASROZ optimization}
KASROZ uses the same physical key grid as ClearFlow (a
$5$-row $4{-}6{-}6{-}6{-}4$ ortho keyboard) but assigns letters
to keys under a different objective. ClearFlow is optimized for
swipe-shape distinctiveness while minimizing trace
length~\citep{clearflowkeyboard}, in the clarity-cost family of
Smith et al.~\citep{smith2015optimizing}. This continues an
older line of keyboard layout-cost engineering that targeted
tap-movement time via Fitts'
law~\citep{zhai2002performance}, swapping movement-time cost
for a shape-ambiguity cost. Shape-distinctiveness costs are a
proxy for decoding ambiguity in template-matching swipe
decoding algorithms. KASROZ replaces this proxy with the
encoder itself.
For each candidate layout, every word in the lexicon is synthesized
into a gesture using the min-jerk path~\citep{flash1985coordination}
through its letter centers. The synthetic swipe is inferenced using the 
encoder, and scored using CTC against the target sequence and candidate layout. The
layout cost is the frequency-weighted sum of these per-word NLLs
plus a Cao-Zhai per-leg duration
term~\citep{cao2007modeling} as an ergonomic counterweight against
the optimizer collapsing frequent letters onto distant parts of
the layout, creating long gesture traces. A batched hill-climb over letter-swap
candidates arrives at the layout in \Cref{fig:kasroz-layout}.
We evaluated more than $800{,}000$ layouts, with KASROZ as the
cost-minimum arrangement.

\paragraph{Colinear letter trigrams hide user intention}
A swipe-keyboard user produces a gesture near or through the positions of the
keys they intend to type. When a target word contains three
consecutive letters that lie nearly colinear on the layout, the
middle letter contributes no visible feature to the swipe. The
curve passes through its key on the way from the previous letter
to the next one whether or not the user meant to select it. From
the encoder's standpoint, that midpoint letter's identity is
under-determined by the swipe. This per-word confusability has
been formalized previously as \emph{word
clarity}~\citep{yi2017wordclarity}. For user experience, this can be
a source of frustration, and maximizing the user's ability to clearly
indicate intention is a layout design choice that can improve usability.

\begin{figure*}[t]
  \centering
  \includegraphics[width=\linewidth]{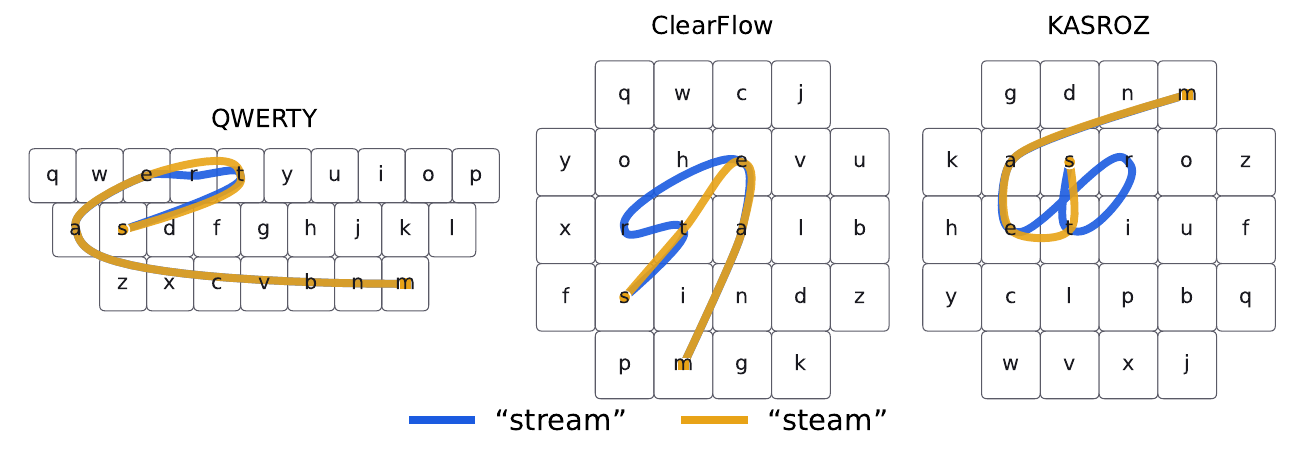}
  \caption{Synthetic min-jerk swipe paths for \texttt{stream}
  (one color) and \texttt{steam} (the other) on three layouts.
  Left: QWERTY. Middle: ClearFlow. Right: KASROZ. The QWERTY
  paths overlap to the point of being a single curve. ClearFlow
  separates the two words but leaves multiple letter trigrams
  near-colinear. KASROZ separates both the words and the
  internal trigrams.}
  \label{fig:kasroz-paths}
\end{figure*}

Swipe geometry is a layout property, not an encoder limitation. When the
encoder can read the curve unambiguously on a layout where confusions
like trigram colinearity are minimized, detection accuracy is naturally improved.
\Cref{fig:kasroz-paths} makes this concrete on the two near-
confusable English words \texttt{stream} and \texttt{steam}.
The two words share \texttt{s}--\texttt{t}--\texttt{e}--\texttt{a}--%
\texttt{m}; \texttt{stream} inserts an \texttt{r} between
\texttt{t} and \texttt{e}. On QWERTY, \texttt{r} sits on the top
row between \texttt{t} and \texttt{e}, so the gesture through it
adds no curvature. The two synthetic paths overlap to the point
of being a single curve.

On ClearFlow, the $4{-}6{-}6{-}6{-}4$
grid places \texttt{r} off the \texttt{t}--\texttt{e} segment and
the two paths separate. But the same arrangement introduces two
new nearly colinear trigrams (\texttt{s}--\texttt{t}--\texttt{e}
and \texttt{e}--\texttt{a}--\texttt{m} in \texttt{steam}), so
those midpoint letters are also under-determined. KASROZ
breaks both trigram configurations because the encoder reported
low letter confidence during the layout search. Although ClearFlow
has successfully added shape uniqueness, it is still sub-optimal for
neural detection.

\begin{figure*}[t]
  \centering
  \includegraphics[width=\linewidth]{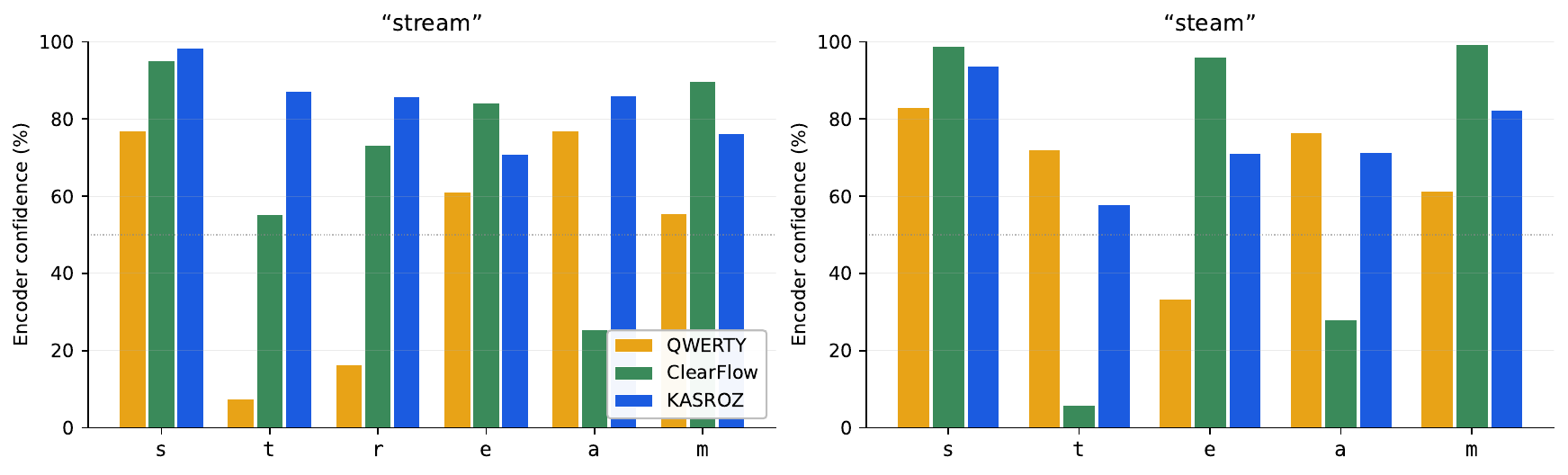}
  \caption{Encoder confidence per letter on synthetic
  \texttt{stream} and \texttt{steam} swipes, by layout.
  Confidence is $e^{-\text{NLL}}$ at the timestep that CTC
  forced-alignment of the target sequence assigned to that
  letter, expressed as a percentage. KASROZ keeps every letter above
  $58\%$; QWERTY drops to $7\%$ on \texttt{t} of \texttt{stream};
  ClearFlow drops to $6\%$ on \texttt{t} of \texttt{steam}.}
  \label{fig:kasroz-decoder-outputs}
\end{figure*}

\Cref{fig:kasroz-decoder-outputs} shows how the encoder reports confidence
for each character in the words. The
colinear-trigram failures on QWERTY and ClearFlow demonstrate
single-letter detection difficulty, with confidences below $10\%$. For these
words, the KASROZ swipe paths give the encoder a clear signal
for each letter.

The user-side consequence is that a QWERTY user who intends
\texttt{stream} and produces a careful \texttt{stream} swipe gesture
may need to dwell over the character or rely on
semantic disambiguation from a context language model to produce
a correct result. The gesture shape alone does not easily
separate \texttt{stream} from \texttt{steam}. KASROZ's
optimization objective addresses this exact phenomenon, measured
letter by letter from the encoder output.

\paragraph{Evaluation on real user swipes}
We collected $2{,}804$ real user swipes against the KASROZ layout
(after the trajectory-quality filter applied in the rest of
the paper) and decode them with the same encoder, beam search,
trie, and scoring constants used for ClearFlow. KASROZ is held
out from training and from scoring tune.

\Cref{tab:kasroz-eval} compares decoding accuracy on real user
swipes across the three English layout variants.
SHARK\textsuperscript{2} rows use the same
EN-tuned constants as in \Cref{tab:main-encoder-only}. Our encoder
rows use the same scoring parameters as the ClearFlow row in
\Cref{tab:main-encoder-only}.

\begin{table}[h]
\centering
\caption{Decoding accuracy on real user swipes across both
$4{-}6{-}6{-}6{-}4$ layouts, plus in-domain QWERTY.}
\label{tab:kasroz-eval}
\begin{tabular}{llccc}
\toprule
Layout & $N$ & Method & Top-1 (\%) & Top-3 (\%) \\
\midrule
QWERTY (in-domain) & $52{,}629$ & SHARK\textsuperscript{2} & $80.05$ & $90.47$ \\
                   &            & Ours                     & $92.94$ & $97.46$ \\
\midrule
ClearFlow          & $11{,}028$ & SHARK\textsuperscript{2} & $92.18$ & $97.23$ \\
                   &            & Ours                     & $96.84$ & $98.98$ \\
\midrule
KASROZ             & $2{,}804$  & SHARK\textsuperscript{2} & $91.19$ & $97.11$ \\
                   &            & Ours                     & \textbf{97.68} & \textbf{99.47} \\
\bottomrule
\end{tabular}
\end{table}

With our encoder, KASROZ is the most accurate layout we measured.
SHARK\textsuperscript{2} performs slightly worse on KASROZ compared
to ClearFlow, as expected. The ClearFlow objective more directly
aligns to the SHARK\textsuperscript{2} algorithm, while KASROZ
directly optimizes the detection quality of our neural model over
the English lexicon. The result
indicates that the synthetic-swipe NLL the layout optimizer
minimizes is a faithful proxy for what the encoder detects on real
user swipes against the same layout. An encoder that reads the layout as a
runtime tensor can both decode layouts designed independently of
it and serve as the cost function for designing new ones.

QWERTY's wide rows of keys produce more ambiguous gestures than
the square-shaped layouts. Despite being the only in-domain
layout for our training data, it performs worse than both
ClearFlow and KASROZ for both neural and template-matching
decoding, which suggests a fundamental layout limit.

\section{Ablations}
\label{sec:ablations}

This section ablates each design choice and measures its effect
on cross-layout transfer. \Cref{ssec:abl-augmentation}
shows that without augmentation, the held-out ClearFlow column
collapses to near-zero (\Cref{tab:abl-augmentation}).
\Cref{ssec:abl-crctc} ablates the fixed-layout decoder training
recipe. The spatial output head of \Cref{ssec:dct} is contrasted
against a learned bilinear-grid alternative in
\Cref{ssec:abl-spatial-head}.

\subsection{Co-augmentation of trajectory and layout}
\label{ssec:abl-augmentation}
This ablation isolates the contribution of co-augmentation to
cross-layout transfer. Every variant trains the same encoder
architecture on English-only \texttt{swipe.futo.org} data with the
same $60$-epoch budget, cosine LR schedule, and seed. Russian and ClearFlow validation
data is held out, so the Russian and ClearFlow columns of
\Cref{tab:abl-augmentation} measure zero-shot transfer to new layouts.

\begin{table}[h]
\centering
\caption{Effect of cumulative co-augmentation stages on encoder-only
top-1 accuracy at the $60$-epoch ablation budget. Every stage is
applied jointly to the trajectory and the layout-key tensor
(\Cref{ssec:augmentation}). Each row uses its own
two-layer tune (\Cref{ssec:setup}). \Cref{tab:main-encoder-only}
extends this same full recipe to $120$ epochs and reaches $83.11\%$
on the same held-out Russian data. Bold marks the column best,
underline the runner-up.}
\label{tab:abl-augmentation}
\begin{tabular}{lccc}
\toprule
Augmentation stages & English & Russian & ClearFlow \\
                    & ($n{=}52{,}629$) & ($n{=}9{,}970$) & ($n{=}11{,}028$) \\
\midrule
Baseline                                                          & \textbf{93.91} & 40.54 & 3.22 \\
\midrule
$+$ rotation                                                      & \underline{93.66} & 60.99 & 83.03 \\
$+$ flips, translation                                            & \underline{93.66} & 70.22 & 93.49 \\
$+$ y-scale, time-reverse                                         & 93.23 & 75.12 & 96.08 \\
$+$ x-scale, shear (full)                                         & 93.25 & \textbf{77.15} & \underline{96.45} \\
\midrule
\multicolumn{4}{l}{\emph{Leave-one-out from the full recipe}} \\
\quad{}$-$ y-scale                                                & 93.59 & 75.18 & 95.29 \\
\quad{}$-$ x-scale                                                & 93.48 & 75.66 & 95.87 \\
\quad{}$-$ flips                                                  & 93.46 & 74.94 & 95.77 \\
\quad{}$-$ time-reverse                                           & 93.34 & 76.77 & 96.35 \\
\quad{}$-$ shear                                                  & 93.30 & \underline{76.80} & \textbf{96.54} \\
\bottomrule
\end{tabular}
\end{table}

Co-augmentation helps cross-layout transfer but not in-layout
accuracy. The English column is flat across rows and trends
slightly \emph{down} from baseline as more stages are added. The augmented encoder gives up a small amount of in-domain QWERTY
accuracy in exchange for accuracy on layouts whose data was not
used for training. The two
held-out columns tell the opposite story. ClearFlow rises from
$3.22\%$ at baseline to above the in-domain QWERTY row. The
Russian column moves in the same direction independently. Under
English-only training and no Russian samples at any stage, the
recipe lifts the held-out Russian row from $40.54\%$ to $77.15\%$
top-$1$. Scoring is tuned on English QWERTY validation data.

The full-recipe row's Russian value sits below the result of
\Cref{tab:main-encoder-only} because the main table extends the
same recipe to $120$ epochs. The model is still
improving at the $60$-epoch ablation budget.

\subsection{Decoder training recipe: CR-CTC $\times$ ranking loss}
\label{ssec:abl-crctc}

\begin{table}[h]
\centering
\caption{Decoder training recipe: $2{\times}2$ of CR-CTC and
ranking loss (LambdaLoss with hard-negative mining),
trained at a $50$-epoch budget, with the two final rows at $100$ epochs.
$\Delta_\text{top1}$ is the
gain over the encoder-only baseline (Wald $95\%$ half-width
$\pm 0.22$ pt at $n{=}52{,}629$). Bold marks the column best,
underline the runner-up.}
\label{tab:abl-crctc}
\begin{tabular}{cccccc}
\toprule
CR-CTC & Ranking & Epochs & Top-1 (\%) & Top-3 (\%) & $\Delta_\text{top1}$ \\
\midrule
$\times$    & $\times$    & 50  & $92.83$ & $97.38$ & $-0.11$ \\
$\times$    & \checkmark  & 50  & $93.27$ & $97.71$ & $+0.33$ \\
\checkmark  & $\times$    & 50  & $93.00$ & $97.50$ & $+0.06$ \\
\checkmark  & \checkmark  & 50  & $93.46$ & \underline{97.84} & \underline{+0.52} \\
\midrule
$\times$    & \checkmark  & 100 & $93.31$ & $97.68$ & $+0.37$ \\
\checkmark  & \checkmark  & 100 & \textbf{93.52} & \textbf{97.85} & \textbf{+0.58} \\
\midrule
\multicolumn{3}{l}{\emph{Encoder-only baseline}} & $92.94$ & $97.46$ & --- \\
\bottomrule
\end{tabular}
\end{table}

Two observations follow from \Cref{tab:abl-crctc}.

\paragraph{Combined recipe is best, additively}
At the $50$-epoch budget the combined recipe outperforms either
component alone, and the two gains are roughly additive. At
$100$ epochs the combined cell is the column best on both top-$1$
and top-$3$. The bare-decoder cell (both terms off) lands
slightly below the encoder-only baseline, showing that the
encoder's DCT head plus length-aware pruning already recovers
most of the word-level accuracy a lexical decoder could add.

\paragraph{CR-CTC and overfitting}
Without CR-CTC the ranking-only recipe overfits steadily after
its val-loss minimum near epoch~$14$. Val top-$1$ peaks at
$93.75\%$ around epoch~$25$ and drifts down by nearly a full
point by the end of the schedule. Adding CR-CTC delays the
val-loss minimum by roughly $20$ epochs and reduces the
post-minimum rise by an order of magnitude. Val top-$1$ ends
the schedule near its peak. CR-CTC makes the training schedule less sensitive to checkpoint
selection, with accuracy staying near-peak across roughly $50$
epochs. This is separate from its modest final-checkpoint top-$1$
gain.

\section{Conclusion}
\label{sec:conclusion}

In this report, we demonstrate a method for producing a
layout-agnostic neural swipe model. The encoder reads the
keyboard as runtime input, and a coordinated augmentation pipeline
teaches it to predict character intent from the gesture itself
rather than from layout-specific features.

Using joint augmentation, a single encoder's zero-shot accuracy
on ClearFlow exceeds the in-domain accuracy on the QWERTY it was
trained on. With a layout-invariant encoder, the choice of
layout becomes an inference-time selection rather than a fixed
input. The approach combines the flexibility of an algorithmic
decoder with the improved accuracy of a neural model, and composes
with downstream components, such as the optional fixed-layout
decoder (\Cref{ssec:decoder}) and the context language model, to
further improve accuracy.

We release the trained models and an MIT-licensed corpus of over
1M donated swipes from more than 12k donor sessions.

\section*{Acknowledgements}

We thank Sameer Suri and Thomas Folbrecht for their contributions to
the \texttt{swipe.futo.org} data collection effort.

\bibliographystyle{unsrtnat}
\bibliography{references}

\appendix
\section{Effect of synthetic Indic data on cross-layout transfer}
\label{app:abl-synth}

To test whether synthetic data from a typologically distant
keyboard family improves the encoder's cross-layout transfer, we
re-train the full-augmentation baseline of
\Cref{ssec:abl-augmentation} with a subset of the IndicSwipe
synthetic corpus~\citep{biju2020indicswipe} added to the training
mix (roughly $170$\,K additional swipes across six Indic scripts,
from the canonical $193{,}658$-swipe / $7$-language release).
Both rows use the same encoder architecture, the same augmentation
pipeline, and a $60$-epoch training budget.

\begin{table}[h]
\centering
\caption{Synthetic Indic data added to training. Same recipe as
the ``$+$ x-scale, shear'' row of \Cref{tab:abl-augmentation}, with
scoring tuned per-row on EN val.}
\label{tab:abl-synthetic-indic}
\begin{tabular}{lcccc}
\toprule
Training data                & EN val (\%) & EN test (\%) & RU val (\%) & CF val (\%) \\
                             & ($n{=}52{,}629$) & ($n{=}48{,}538$) & ($n{=}9{,}970$) & ($n{=}11{,}028$) \\
\midrule
EN only                       & $93.25$ & $93.13$ & $77.15$ & $96.45$ \\
EN $+$ IndicSwipe synthetic   & $93.39$ & $93.22$ & $76.68$ & $96.32$ \\
\bottomrule
\end{tabular}
\end{table}

We interpret the synthetic IndicSwipe data as too easy at training
time to contribute meaningful gradient. The trajectories are
generated by a parametric model of the target word's key sequence
rather than recorded from users, so they lack the motor noise,
hesitation, and curvature mismatch that real swipes exhibit. As
direct evidence, the same encoder reaches $96.5\%$ greedy-CTC
top-1 on the held-out synthetic Tamil validation split, i.e.\ the
synthetic distribution is trivial to fit, even without a lexicon. This result
should not be read as evidence against multi-layout training, only
against the efficacy of synthetic data as a substitute for real data.

\section{Mobile deployment details}
\label{app:mobile-deployment}

This appendix collects the deployment-design and on-device profile
material referenced from \Cref{sec:method} and \Cref{sec:experiments}.

\paragraph{Layout at runtime}
The exported encoder takes three tensor inputs:
\begin{align*}
  \texttt{features}    &: [1,\, 2,\, T_{\text{in}}] \quad \text{(raw $(x,y)$ trajectory, $T_{\text{in}}{=}64$)}, \\
  \texttt{layout\_keys} &: [1,\, K_{\max},\, 2] \quad \text{(per-key $(c_x, c_y)$, zero-padded)}, \\
  \texttt{layout\_mask} &: [1,\, K_{\max}] \quad \text{(bool mask, True for real keys).}
\end{align*}
We fix $K_{\max} = 64$ at export time, sized to accommodate
Indic-scale alphabets (e.g., Devanagari) with headroom. Tensor
shapes for every stage of the forward pass are in
\Cref{tab:encoder-tensor-shapes}, where $B$ is the batch size,
$T_{\text{in}} = 64$ and $T_{\text{out}} = 32$ are the time-axis
lengths before and after the adapter, and $H$ is the backbone
hidden width.

\begin{table}[h]
\centering
\caption{Tensor shapes through the deployed encoder forward pass.}
\label{tab:encoder-tensor-shapes}
\begin{tabular}{lll}
\toprule
Tensor                         & Shape              & Source \\
\midrule
input trajectory               & $[B,\, 2,\, T_{\text{in}}]$    & raw $(x, y)$ \\
backbone output                & $[B,\, T_{\text{in}},\, H]$    & TCN \\
adapter output                 & $[B,\, T_{\text{out}},\, H]$   & 2$\times$ adapter \\
coefficients $\vect{c}_t$      & $[B,\, T_{\text{out}},\, 64]$  & coefficient projection \\
gate $\lambdat$                & $[B,\, T_{\text{out}}]$        & $\sigma(\cdot)$ projection \\
layout-key tensor              & $[B,\, K_{\max},\, 2]$         & runtime input \\
basis $\dctbasis$              & $[B,\, K_{\max},\, 64]$        & \Cref{eq:dct-basis} \\
key logits $\vect{z}_t$    & $[B,\, T_{\text{out}},\, K_{\max}]$ & $\vect{c}_t\, \dctbasis^{\top}$ \\
log-emissions over $K{+}1$ classes & $[B,\, T_{\text{out}},\, K_{\max}{+}1]$ & \Cref{app:abl-lambda} \\
\bottomrule
\end{tabular}
\end{table}

\paragraph{Inference modes}
The exported binaries support two inference modes which differ only
in which files are loaded at runtime. \emph{Encoder-only} runs beam
search directly on the encoder's $\lambdat$-gated log-emissions
over the layout's lexicon trie. \emph{Encoder plus decoder} runs the
fixed-layout decoder \texttt{.pte} on the encoder's output features
and then beams over the corrected log-emissions. The zero-init
residual skip of \Cref{fig:decoder-skip} makes this mode a strict
superset of encoder-only at the model level.

\paragraph{On-device profile}
\Cref{tab:device-profile} reports on-disk size, parameter count,
and forward latency on a Google Pixel~4 (Snapdragon~855, ARM~v8a)
under single-threaded ExecuTorch with XNNPACK delegation. Latency
is pinned to the four Cortex-A76 performance cores and to the four
Cortex-A55 efficiency cores over $500$ runs after a $50$-run
warmup. $p50$ is reported because $p99$ is dominated by OS
scheduling noise. End-to-end is the full pipeline: resample,
encoder, optional decoder, trie-constrained beam search of width
$100$.

\begin{table}[h]
\centering
\caption{On-device profile on a Google Pixel~4 (Snapdragon~855,
ARM~v8a, single-threaded). The encoder \texttt{.pte} is shared
across all layouts and exported in mixed precision (fp16 backbone,
fp32 spatial head). The optional decoder \texttt{.pte} is
English-only and exported in fp16.}
\label{tab:device-profile}
\begin{tabular}{lcc}
\toprule
                                & Encoder \texttt{.pte} & Per-layout decoder \texttt{.pte} \\
                                & (one for all layouts) & (English, optional)              \\
\midrule
Parameters                      & $635$K   & $304$K   \\
File size, fp32                 & 2.6\,MB  & 1.2\,MB  \\
File size, fp16                 & 2.5\,MB  & 0.65\,MB \\
Forward latency, A76 $p50$ (ms) & $1.54$   & $0.45$   \\
Forward latency, A55 $p50$ (ms) & $7.56$   & $2.83$   \\
\midrule
End-to-end $p50$ (ms / swipe), A76 cores
                                & \multicolumn{2}{c}{$2.13$ (encoder only),\ \ $2.78$ (encoder + decoder)} \\
End-to-end $p50$ (ms / swipe), A55 cores
                                & \multicolumn{2}{c}{$10.1$ (encoder only),\ \ $14.0$ (encoder + decoder)} \\
\bottomrule
\end{tabular}
\end{table}

\section{Augmentation stages}
\label{app:augmentation-stages}

The augmentation pipeline of \Cref{ssec:augmentation} is a
composition of seven stages, in order:
\begin{enumerate}
  \item \textbf{Y-scale.} A per-sample scale $s_y \sim
    \mathcal{U}(0.75, 1.0)$ contracts the $y$ axis around
    $y{=}0.5$. Skipped per-sample for layouts with more than three
    rows (Indic scripts) to avoid violating row geometry.
  \item \textbf{X-scale.} An independent per-sample scale
    $s_x \sim \mathcal{U}(0.85, 1.0)$ contracts the $x$ axis
    around $x{=}0.5$. Simulates layouts whose keys-per-row count
    differs from the training layout, producing narrower or wider
    key cells.
  \item \textbf{Shear.} Two independent per-sample shear factors
    $s_{xy}, s_{yx} \sim \mathcal{U}(-0.05, 0.05)$ apply a small
    affine skew: $x' = x + s_{xy}(y - 0.5)$ then
    $y' = y + s_{yx}(x' - 0.5)$. Breaks the prior that keys lie on
    a strictly orthogonal grid.
  \item \textbf{Flips.} Independent Bernoulli$(0.5)$ flips along
    each axis.
  \item \textbf{Rotation.} A per-sample angle
    $\theta \sim \mathcal{U}[0, 2\pi)$ rotates trajectory and
    keys around the trajectory's centroid. If the rotated content
    would overflow the unit square the rotation is rejected for
    that sample (the pre-rotation state is kept).
  \item \textbf{Translation.} A bounded shift moves the combined
    bounding box of the trajectory and the masked key positions to
    a random valid origin inside $[0,1]^2$.
  \item \textbf{Time reversal.} With probability $0.1$ the temporal
    axis of the trajectory is reversed, and the target word is
    also reversed so the CTC label sequence stays aligned.
\end{enumerate}
Geometric stages 1--6 are applied identically to the trajectory
tensor $[B,\, 2,\, T]$ and to the training-time layout-key tensor
$[B,\, K,\, 4]$, where the two extra columns are the key
half-radii $(r_x, r_y)$ used only to keep the augmented keyboard
geometrically consistent. The radii are scaled along with the
corresponding axis so that each augmented key retains its physical
area on the augmented keyboard.

\section{Training details}
\label{app:training-details}

\Cref{tab:training-details} lists the optimizer schedule and
regularization values for the encoder of \Cref{ssec:dct} and the
fixed-layout decoder of \Cref{ssec:decoder}.

\begin{table}[h]
\centering
\caption{Training hyperparameters for the production encoder and
the fixed-layout English decoder. Loss weights are absolute (not
normalized to sum to 1).}
\label{tab:training-details}
\begin{tabular}{lll}
\toprule
                              & Encoder                 & Decoder                  \\
\midrule
Optimizer                     & AdamW                   & Lion~\citep{chen2023lion} \\
Base LR                       & $1{\times}10^{-3}$      & $3{\times}10^{-4}$       \\
LR schedule                   & cosine to $2{\times}10^{-5}$ & constant after warmup \\
Warmup                        & $5\%$ of steps          & $3$ epochs linear        \\
Betas                         & $(0.9, 0.999)$          & $(0.9, 0.98)$            \\
Weight decay                  & $1{\times}10^{-4}$      & $3{\times}10^{-3}$       \\
Gradient norm clip            & $1.0$                   & $1.0$                    \\
Batch size                    & $1024$                  & $2048$                   \\
Epochs                        & $120$                   & $100$                    \\
Backbone dropout              & $0.1$                   & $0.05$                   \\
EMA decay                     & none                    & $0.999$                  \\
\midrule
Training loss                 & $1.0\,\mathcal{L}_{\text{CTC}} + 0.05\,\mathcal{L}_{\text{emit}}$ & $0.3\,\mathcal{L}_{\text{CTC}} + 5.0\,\mathcal{L}_{\text{rank}} + 0.1\,\mathcal{L}_{\text{CR}}$ \\
CR-CTC noise $\sigma$         & ---                     & $0.10$                   \\
Rank gate threshold           & ---                     & val-CTC $< 0.205$        \\
LambdaLoss $(\mu, \sigma)$    & ---                     & $(10, 1.0)$              \\
Hard-negative pool size       & ---                     & up to $128$ per word     \\
\midrule
Scoring head optimizer        & ---                     & SGD, momentum $0.9$      \\
Scoring head LR               & ---                     & $5{\times}10^{-3} \to 10^{-3}$ cosine \\
Scoring init $(\gamma, \lambda_\text{f}, \beta)$ & ---     & $(0.30, 0.025, 1.80)$, $\gamma$ frozen \\
\bottomrule
\end{tabular}
\end{table}

\section{SHARK\textsuperscript{2} baseline: tuned constants}
\label{app:shark2}

The SHARK\textsuperscript{2}~\citep{kristensson2004shark2} baseline
in \Cref{tab:main-encoder-only} is re-implemented from \S3 of the
original paper, and its free constants are tuned on EN val under
the same Optuna protocol as the encoder scoring tune
(\Cref{ssec:setup}).

\begin{table}[h]
\centering
\caption{SHARK\textsuperscript{2} constants after Optuna tuning
on EN val.}
\label{tab:shark2-tuned}
\begin{tabular}{ll}
\toprule
Constant & Value \\
\midrule
$\theta_{\text{prune}}$ (endpoint pruning) & $0.20$ (fixed) \\
Shape weight                                & $1.0$ (fixed) \\
$\lambda_L$ (location weight)               & $2.99$ \\
$\lambda_f$ (frequency-prior weight)        & $0.40$ \\
\bottomrule
\end{tabular}
\end{table}

\section{Scoring-term ablation across layouts}
\label{app:abl-scoring-terms}

\Cref{tab:abl-scoring-terms} reports the contribution of each
subset of scoring terms in \Cref{eq:scoring} on the encoder-only
path. Each row tunes only the active terms on English val by
maximizing tt1mm5 (Optuna TPE, $2{,}000$ trials per row) at the
main-table pruning, with inactive parameters clamped to zero.
``Raw CTC'' picks the highest-CTC candidate from the beam.

On English and ClearFlow the frequency prior $\lambda_{\text{f}}$
carries most of the improvement over raw CTC, and the
$\lambda_{\text{f}} + \beta$ subset recovers the full three-term
tune within noise. On Russian the pattern inverts: $\lambda_{\text{f}}$
alone falls below raw CTC, $\gamma$ is the strongest single
term, and the three-term tune lands only fractionally above the
best two-term subset (inside the RU CI). The scoring formula is
calibrated to lexicon and language specifics, so when a small
validation pool for a target language is available, retuning
scoring on that pool is a cheap step toward improving in-language
accuracy without retraining the encoder.

\begin{table}[h]
\centering
\caption{Top-1 contribution of each subset of scoring terms in
\Cref{eq:scoring} on the encoder-only path. Each row tunes only the
active terms on English val (Optuna TPE, $2{,}000$ trials), with
inactive terms clamped to zero. Pruning is held at the main-table
$(\gamma_p, \beta_p) = (0.186, 1.139)$ across all rows. Russian and
ClearFlow are held out from the tune. Bold marks the column max,
underline the runner-up.}
\label{tab:abl-scoring-terms}
\begin{tabular}{lcccc}
\toprule
Active terms                        & EN val (\%) & EN test (\%) & RU val (\%) & CF val (\%) \\
\midrule
Raw CTC                             & $83.27$ & $82.05$ & $83.15$ & $88.69$ \\
\midrule
$\gamma$                            & $86.24$ & $85.30$ & \textbf{85.93} & $92.52$ \\
$\lambda_{\text{f}}$                & $89.09$ & $88.43$ & $80.60$ & $92.47$ \\
$\beta$                             & $86.31$ & $85.38$ & $85.61$ & $90.84$ \\
\midrule
$\gamma, \lambda_{\text{f}}$        & $91.36$ & $90.95$ & $83.30$ & $94.99$ \\
$\gamma, \beta$                     & $86.42$ & $85.60$ & \underline{85.73} & $92.23$ \\
$\lambda_{\text{f}}, \beta$         & \textbf{92.98} & \textbf{92.62} & $82.73$ & \textbf{96.83} \\
\midrule
$\gamma, \lambda_{\text{f}}, \beta$ & \underline{92.95} & \underline{92.56} & $83.44$ & \underline{96.82} \\
\bottomrule
\end{tabular}
\end{table}

\section{Language dependence of the scoring tune}
\label{app:abl-scoring-language}

The $83$\% top-1 the scoring achieves on held-out Russian
raises an obvious question: how much of that gap is the encoder
failing to generalize across alphabets, and how much is the
scoring formula being calibrated to English's word-frequency and
word-length distribution?

\Cref{tab:abl-scoring-language} answers this directly. The same
EN-only encoder is held fixed across all three rows. Only the val
pool used for the two-layer tune (outer pruning on recall@$K$ and inner scoring on
tt1mm5) varies. Each row's pruning and scoring share the same
scope so that an EN-only-tuned pruning is never paired with an
RU-tuned scoring or vice versa. ClearFlow is held out from every
row's tune and reported as a fourth column.

\begin{table}[h]
\centering
\caption{Language dependence of the two-layer tune. Each row is a
full two-layer tune: outer pruning fit against recall@$100$ on the
scope's val pool, inner scoring fit against tt1mm5 on the same
scope. ClearFlow is held out from all three tunes. Bold marks the
column max.}
\label{tab:abl-scoring-language}
\begin{tabular}{lccccc cccc}
\toprule
                                  & \multicolumn{2}{c}{Pruning}            & \multicolumn{3}{c}{Scoring}                                & \multicolumn{4}{c}{Top-1 (\%)} \\
\cmidrule(lr){2-3} \cmidrule(lr){4-6} \cmidrule(lr){7-10}
Tune scope                        & $\gamma_p$ & $\beta_p$ & $\gamma$ & $\lambda_{\text{f}}$ & $\beta$ & EN val & EN test & RU val & CF val \\
\midrule
EN val only           & $0.186$ & $1.139$ & $0.105$ & $0.050$ & $2.488$ & \textbf{92.94} & \textbf{92.54} & $83.11$ & \textbf{96.84} \\
RU val only           & $0.314$ & $1.032$ & $0.500$ & $0.008$ & $0.590$ & $90.23$ & $89.48$ & \textbf{86.59} & $95.20$ \\
EN + RU joint         & $0.072$ & $2.049$ & $0.700$ & $0.011$ & $0.390$ & $91.70$ & $91.17$ & $86.09$ & $96.09$ \\
\bottomrule
\end{tabular}
\end{table}

Retuning scoring on Russian val alone recovers most of the
English-Russian gap on Russian at the cost of several points on
English. The calibration of the scoring formula is
language-specific, and can be fit from a val pool at least an
order of magnitude smaller than the training set. Pre-normalizing
the language dictionary frequencies (z-score or rank) does not
close this gap, suggesting scale mismatch is not the primary cause.

The EN+RU joint tune lands close to the EN-only tune on English
while recovering nearly all the Russian gain over the EN-only
tune. Adding one further language to the tune-time val mix is a
cheap way to broaden deployment scoring without sacrificing
in-domain accuracy.

ClearFlow top-1 is roughly insensitive to which language is
tuned. The swipe-optimized layout produces sharper encoder
emissions than QWERTY, which makes its top-1 less responsive to
changes in the scoring parameters.

\section{Beam-width sensitivity}
\label{app:abl-beam}

\begin{table}[h]
\centering
\caption{Encoder-only top-$k$ as a function of beam width. Same
production encoder, trie, scoring, and pruning as
\Cref{tab:main-encoder-only}. Bold marks the column best, underline
the runner-up.}
\label{tab:abl-beam-width}
\setlength{\tabcolsep}{4pt}
\begin{tabular}{lccccccccc}
\toprule
       & \multicolumn{3}{c}{EN val (FUTO)} & \multicolumn{3}{c}{RU val (Yandex)} & \multicolumn{3}{c}{CF val (held)} \\
\cmidrule(lr){2-4} \cmidrule(lr){5-7} \cmidrule(lr){8-10}
Beam   & Top-1 & Top-3 & Top-10 & Top-1 & Top-3 & Top-10 & Top-1 & Top-3 & Top-10 \\
\midrule
$10$   & $88.74$ & $91.88$ & $92.16$ & $82.13$ & $89.39$ & $91.12$ & $95.49$ & $97.25$ & $97.36$ \\
$25$   & $91.42$ & $95.40$ & $96.05$ & $82.84$ & $90.55$ & $94.03$ & $96.30$ & $98.32$ & $98.66$ \\
$50$   & $92.48$ & $96.78$ & $97.71$ & $83.02$ & $91.09$ & $94.97$ & $96.71$ & $98.82$ & $99.27$ \\
$100$  & $92.94$ & $97.46$ & $98.60$ & \textbf{83.11} & $91.26$ & $95.42$ & $96.84$ & $98.98$ & $99.48$ \\
$200$  & \underline{93.13} & \underline{97.74} & \underline{99.02} & \textbf{83.11} & \textbf{91.48} & \underline{95.79} & \underline{96.91} & \underline{99.07} & \underline{99.59} \\
$400$  & \textbf{93.20} & \textbf{97.85} & \textbf{99.19} & \underline{83.08} & \underline{91.42} & \textbf{95.86} & \textbf{96.94} & \textbf{99.12} & \textbf{99.66} \\
\bottomrule
\end{tabular}
\end{table}

Top-$1$ on all three layouts saturates by width $100$. English
gains a fraction of a point at widths $200$ and $400$, while
Russian and ClearFlow move within their respective CIs across the
same range. By preventing long-prefix hypotheses from being out-competed by
shallower candidates with less accumulated CTC cost, length-aware
pruning lets smaller beam widths reach near-peak accuracy.

\section{Spatial output head: spectral basis, learned grid, and disc support}
\label{ssec:abl-spatial-head}
\label{app:disc-head}

The fixed cosine basis of \Cref{ssec:dct} is not the only way to
build a layout-agnostic spatial head. Any head that takes the
key centroids $(x, y)$ position at evaluation time and emits key
logits is layout-agnostic by construction. This appendix contrasts
the fixed-basis DCT head with a same-shape learned bilinear grid,
characterizes the spectral compactness of the trained head, and
reports a quarter-disc-supported variant that exploits that
compactness. The grid replaces the cosine basis with an $N \times N$
table of learned coefficients, interpolated at the runtime key
positions. The DCT head is smooth in $(x, y)$ and evaluates in
closed form. The grid head is $C^0$-continuous at cell boundaries
and learns its own spatial parameters. Every variant trains the
same backbone on English QWERTY-only \texttt{swipe.futo.org} with
the co-augmentation pipeline of \Cref{ssec:abl-augmentation}.

\begin{table}[h]
\centering
\caption{Encoder accuracy as a function of spatial-head family and
resolution $N$. Trie-beam columns use beam width $100$ search over
the lexicon trie. All variants are trained English-only
with the recipe of \Cref{ssec:abl-augmentation}. Russian and ClearFlow
are held out from both training and tuning. Each row uses its own
two-layer tune on English val. Bold marks the column best, underline
the runner-up.}
\label{tab:abl-spatial-head}
\begin{tabular}{llcccc}
\toprule
     &      & Head      & EN val (\%) & RU val (\%) & CF val (\%) \\
Head & $N$  & params    & ($n{=}52{,}629$) & ($n{=}9{,}970$) & ($n{=}11{,}028$) \\
\midrule
DCT  & $2$  & $1{,}293$  & $64.37$ & $20.45$ & $37.84$ \\
DCT  & $3$  & $2{,}588$  & $92.87$ & $76.53$ & $96.04$ \\
DCT  & $4$  & $4{,}401$  & $93.06$ & $76.67$ & $96.18$ \\
DCT  & $8$  & $16{,}833$ & $93.25$ & $77.15$ & \textbf{96.48} \\
DCT  & $12$ & $37{,}553$ & \underline{93.37} & \underline{77.33} & \underline{96.48} \\
DCT  & $16$ & $66{,}561$ & $93.26$ & \textbf{78.07} & $96.24$ \\
\midrule
Grid & $12$ & $37{,}553$ & \textbf{93.43} & $77.19$ & $96.31$ \\
Grid & $16$ & $66{,}561$ & $93.23$ & $76.58$ & $96.35$ \\
\bottomrule
\end{tabular}
\end{table}

Three observations follow from \Cref{tab:abl-spatial-head}.

\paragraph{Accuracy plateau}
The minimum $N=2$ basis collapses on all three columns. Moving
to $N=3$ recovers nearly the full accuracy of the largest head
tested. From $N=3$ through $N=16$ each column moves within
roughly a one-point band. Accuracy is set by the lowest few DCT
coefficients. Adding capacity beyond that produces only marginal
changes in each column's best result.

\paragraph{DCT vs grid}
At matched resolution the two heads tie on in-domain English
(both deltas inside the EN CI). On the held-out columns the DCT
leads, with its Russian advantage at $N{=}16$ exceeding the CI.
The DCT's smooth basis extrapolates to unseen layouts better than
the grid's piecewise cells. The DCT also carries no learned spatial parameters, so it
matches or exceeds the grid at lower deployment cost.

\section{Blank handling and emission-count penalty}
\label{app:abl-lambda}

This appendix specifies the blank-handling factorization and the
emission-count regularizer used during encoder training.

\subsection{Blank-gate factorization}
\label{ssec:lambda}

\paragraph{Adopted factorization}
Let $\vect{z}_t \in \R^{K}$ be the per-key logits from
\Cref{eq:dct-lookup} and $\lambdat \in [0,1]$ a sigmoid scalar
emitted by an independent head. Following \citet{chao2020varctc}
(Eq.~6--7) we factor the per-timestep emission distribution over
$K{+}1$ classes (characters $1..K$ followed by blank) as
\begin{equation}
\log p_{t,k}
  \;=\;
  \begin{cases}
    \log \sigma_k(\vect{z}_t) \;+\; \log \lambdat & k = 1, \dots, K, \\
    \log(1 - \lambdat) & k = \text{blank},
  \end{cases}
  \label{eq:lambda-emission}
\end{equation}
where $\sigma_k$ is softmax over the key axis. The CTC loss is
computed on the resulting log-emission distribution. The same
factorization appears as the prior head in \citet{chao2020varctc}
and as the per-pair sigmoid blank gate $b_{t,u}$
in \citet{variani2020hat}.

\paragraph{Implementation}
The gate is a \texttt{nn.Linear(hidden, 1)} projection followed by
a sigmoid, sharing the backbone hidden state with the coefficient
projection but otherwise independent. The bias is zero-initialized
so $\lambdat = \sigma(0) = 0.5$ uniformly at step zero. We adopt
this factorization for two reasons. First, $\lambdat$ is a
per-timestep scalar that the fixed-layout decoder (\Cref{ssec:decoder})
consumes independently of which key was predicted. Second, the
emission-count penalty below is a sum of $\lambdat$ over
timesteps. The equivalent constraint under the $(K{+}1)$-way
softmax would route through the shared denominator.

\subsection{Emission-count penalty}
\label{ssec:emission}

For target word length $\ell_{\text{tgt}}$ and predicted gate sum
$\sum_t \lambdat$, we add a one-sided quadratic penalty
\begin{equation}
  \mathcal{L}_{\text{emit}}
  \;=\;
  \alpha \cdot \max\!\bigl(0,\,
       \ell_{\text{tgt}} - \textstyle\sum_t \lambdat\bigr)^{2},
  \quad \alpha = 0.05,
\label{eq:emission-penalty}
\end{equation}
to the standard CTC loss. The penalty activates only when the
model under-emits and has zero gradient once enough mass is
allocated. Over-emission is left to the CTC loss. Production uses
$\alpha = 0.05$.

\end{document}